\documentstyle[mprocl]{article}

\def\be{\begin{equation}}
\def\ee{\end{equation}}
\def\bea{\begin{eqnarray}}
\def\eea{\end{eqnarray}}

\begin{document}

\title{Vortex in Chiral Superconducting State}

\author{J. Goryo}

\address{Department of Physics, Hokkaido University,
\\Sapporo, 060-0810 Japan \\E-mail: goryo@particle.sci.hokudai.ac.jp}


\maketitle\abstracts{
We have investigated the vortex in chiral superconductors, 
especially in $p$-wave case.  
In chiral superconductors the Cooper pair has orbital angular momentum hence   
$U(1)$, parity (P) and time reversal symmetry (T) are broken simultaneously. 
We have found that the vortex has fractional charge and 
fractional angular momentum which comes from P- and T-violation. 
The fractionalization of the angular momentum suggests that the vortex could 
be anyon which obeys the fractional statistics.      
We have also pointed out that the electric field is induced near the 
vortex core and non-trivial electromagnetic phenomena are expected to occur. 
}

In unconventional superconductivity, it has been pointed out  
the possibility of the existence of chiral (P- and T-violating) 
superconducting states, in which the condensation of Cooper pairs 
that have orbital angular momentum occurs. \cite{Sigrist-Ueda} 
Recently, it is argued the realization of the chiral 
$p$-wave ($p_{x}+ip_{y}$-wave) superconductivity 
in $Sr_{2}RuO_{4}$.\cite{Maeno} 
Chiral $d$-wave ($d_{x^{2}-y^{2}} + i d_{xy}$-wave) superconductivity has been 
proposed in High-T$_{c}$ superconductors.\cite{Laughlin2} 
It is important to investigate P- and T-violating phenomena to probe 
the chiral superconductivity. In the present paper, we investigate the vortex 
in chiral superconductors, especially in $p$-wave case, and show that 
it has fractional charge and fractional angular momentum 
as the consequence of P- and T-violation.\cite{Goryo,A-H-C-S} 

$Sr_{2}RuO_{4}$ has a layered 
perovskite crystal structure without cupper and has low transition temperature 
$T_{c} \simeq 1.5[K]$, and is a strongly correlated 2-dimensional Fermi liquid 
in normal state.
Rice and Sigrist proposed that pairing symmetry of orbital part 
is chiral $p$-wave, which has the same orbital symmetry of 
superfluid $^{3}$He-A.\cite{He-A text}
The proposal is consistent with the discovery of internal magnetic field 
in the superconducting phase by $\mu$SR measurement 
and the experiment of $^{17}$O-NMR Knight shift.\cite{muSR-NMR} 
In the chiral $p$-wave superconductor, Cooper pair has orbital angular 
momentum $l_{z}=1$, where $z$ is perpendicular to 
the superconducting plane, {\it i.e.}, the same direction as   
the c axis of the crystal. Therefore, local $U(1)$ gauge symmetry, 
P and T are spontaneously broken 
by the same order parameter.

The above situation is analogous to the quantum Hall system (QHS), 
which is the 2-dimensional electron system in 
an external magnetic field.
In QHS the external magnetic field violates P and T,    
but the explicit local $U(1)$ gauge symmetry is preserved.  
It was shown that the Chern-Simons term is 
induced in the effective action of the gauge field 
when Fermion is integrated out.\cite{Ishikawa} 
The Chern-Simons term is a P- and T-odd bi-linear form of 
the gauge fields and has one derivative. Therefore, this term 
plays important roles in low energy and long distance physics 
and causes P- and T-violating electromagnetic phenomena.  
The coefficient of the induced Chern-Simons term is 
quantized exactly as (integer)$\times$ $\frac{e^{2}}{2 \pi}$. 
$U(1)$ gauge invariance guarantees the exact quantization. 
Since the coefficient of the term becomes 
the Hall conductance, the integer quantum Hall effect was  
explained from gauge invariance.\cite{Ishikawa,TKNN}
     
It was shown that the Chern-Simons term is also
induced in the Ginzburg-Landau action of chiral $p$-wave 
superconductors by integrating out Fermion and $U(1)$ Goldstone mode .
\cite{Goryo-Ishikawa} 
The term has a form written as 
\begin{equation}   
\int d^{3} x \frac{\sigma_{xy}}{2} 
\varepsilon^{ij} (A_{0}^{\rm T} \partial_{i} A_{j} 
+ A_{i} \partial_{j} A_{0}^{\rm T}).    
\label{C-S} 
\end{equation} 
$A_{0}^{\rm T}$ is the transversal component of $A_{0}$ written as 
$A_{0}^{\rm T} = A_{0} - 
\frac{\partial_{0}}{\partial_{0}^{2} - c_{s}^{2}{\bf \partial}^{2}}
(\partial_{0} A_{0} - c_{s}^{2}{\bf \partial} \cdot {\bf A})
$.
In this case, the Chern-Simons term does not have exactly quantized 
coefficient $\sigma_{xy}$ written as 
\begin{equation}   
\sigma_{xy}=
\frac{i}{2!} \varepsilon^{ij} 
\frac{\partial}{\partial q^{j}} \pi_{0i}(q) |_{q=0}
=\frac{e^{2}}{8}\int \frac{d^{2} p}{(2 \pi)^{2}} 
\frac{tr[\vec{g} \cdot ({\bf \partial}\vec{g} \times {\bf \partial}\vec{g})
-g_{3} ({\bf \partial}\vec{g} \times {\bf \partial}\vec{g})_{3}]}
{tr[\frac{1}{2}{\vec g} \cdot {\vec g}]^{\frac{3}{2}}}, \label{Hallcond3}
\end{equation}
here $\pi_{0i}(q)$ is the density-current correlation function, 
$tr$ means trace about {\it real} spin indices and 
${\bf \partial}=\frac{\partial}{\partial {\bf p}}$. 
$\vec{g}({\bf p})$ is related to Fermion propagator 
$G(p_{0},{\bf p})=( p_{0} - \vec{g}({\bf p}) \cdot \vec{\tau} )^{-1}$ 
in Bogoliubov-Nambu Fermion representation   
$\Psi_{\alpha}=\left(\begin{array}{c} 
\psi \\
\psi^{\dagger}
\end{array}\right) 
$ with isospin $\alpha=1,2$, and 
 $\vec{\tau}$ is $2\times2$ Pauli matrices with isospin indices.  
The first term in Eq.(\ref{Hallcond3}) is a topological 
invariant. Volovik have argued this type of topological invariant. 
\cite{Volovik} 
The second term is not a topological invariant. 
The second term appears because $U(1)$ gauge symmetry is 
spontaneously broken.\cite{Goryo-Ishikawa}
Typical $\vec{g}$ for chiral $p$-wave superconductor with 
1 band circular Fermi surface is written as 
\begin{equation}
{\vec{g}}({\bf {p}})= 
\left(\begin{array}{c} 
|\eta_{0}|\hat{p_{x}} i \sigma_{3} \sigma_{2} \\ 
-|\eta_{0}|\hat{p_{y}} i \sigma_{3} \sigma_{2} \\
\frac{{\bf p}^{2}}{2 m} - \epsilon_{\rm F} 
\end{array} \right),  
\label{g-vec}
\end{equation}
$\vec{\sigma}$ is $2 \times 2$ Pauli matrices with real spin indices. 
By substituting Eq. (\ref{g-vec}) into Eq. (\ref{Hallcond3}), we 
obtain the value 
\begin{equation}
\sigma_{xy}=\frac{e^{2}}{4 \pi}, 
\label{sigma_xy}
\end{equation}
which coincides with the fine structure constant. 
In this case, 
the non-topological term in Eq. (\ref{Hallcond3}) vanishes 
{\it accidentally}. In general, 
the non-topological term does not vanish. Actually, if we calculate 
$\sigma_{xy}$ in the tight binding scheme, the non-topological term 
has negligibly small value and $\sigma_{xy}$ is approximately equal 
to the fine structure constant.\cite{Goryo-Ishikawa} The existence of the 
Chern-Simons term suggests that a Hall effect without magnetic field 
occurs.\cite{Goryo-Ishikawa} \cite{Volovik}

Next we consider a vortex in chiral $p$-wave superconductor. 
Except for the Chern-Simons term, 
the quite general phenomenological Ginzburg-Landau free energy 
of the chiral $p$-wave superconductor has been proposed 
by Sigrist and Ueda\cite{Sigrist-Ueda} with tetragonal symmetry 
which corresponds to the layered perovskite structure of the crystal lattice.  
We combine both of them and investigate the vortex solution. 
The vortex charge is defined by the Gauss law which is modified by 
the Chern-Simons term and it is calculated as 
\begin{equation}
Q=
\int d^{2}x \left\{{\bf \nabla} \cdot {\bf E} - 
\sigma_{xy} {\rm B} \right\} 
=- \sigma_{xy} \frac{2 n \pi}{2 e} = - \frac{n e}{4}, 
\label{fractional-charge}\end{equation}
because the electric field is screened. $n=0,\pm 1,\pm 2 \cdot \cdot \cdot$ 
is the vorticity. 
We can see the flux $\frac{2 n \pi}{2 e}$ is attached to the charge,  
{\it i.e.} the vortex is an object such as flux-charge composite and 
analogous to Laughlin's quasiparticle in fractional quantum Hall state. 
It was argued by many authors that the change in the density of electrons 
in the vortex core due to the spatial dependence of the orderparameter 
also gives the charge of the vortex\cite{charging} 
and we call it ``the regular charge''. 
Recently, Volovik pointed out that the occupation of 
the zero-energy bound states of electrons, which exist in 
the vortex core of the chiral p-wave superconductor,\cite{Kopnin-Salomaa} 
could contributes the charge of the vortex.
\cite{Volovik-0-mode,Jackiw-Rebbi}  
We call it ``zero-mode charge''.  
It is important how to distinguish these two charges and 
the fractional charge Eq. (\ref{fractional-charge}) that comes from 
the Chern-Simons term.  
It can be done, even in the case that the regular charge or 
the zero-mode charge are of order $e$, by comparing the vortex 
charges of vorticities $n=1$ and $n=-1$ because the fractional 
charge Eq. (\ref{fractional-charge}) 
changes its sign but other two charges do not depend on 
the sign of vorticity.  
The definition of the vortex angular momentum is  
\begin{equation} 
J=\int d^{2}x {\bf r} \times {\bf {\cal {P}}}. 
\label{angular momentum}   
\end{equation}  
${\cal {P}}$ is the momentum density which is defined as the 
generator of the translation. We find $J$ is a integral  
of the total derivative terms of the gauge fields and 
proportional to $\sigma_{xy}$. \cite{Goryo} $J$ takes 
the fractional value such as 
\begin{eqnarray} 
J=-\frac{n^{2}}{16}.
\label{fractional value}
\end{eqnarray}  
Since the charge and the angular momentum are proportional to $\sigma_{xy}$,  
we see that the Chern-Simons term creates 
these non-zero fractional values. 
We have found that these values are topological and depends only on asymptotic 
behavior of the gauge fields, hence they do not depend on the form of 
Ansatz and also do not depend on the phenomenological parameters 
in Ginzburg-Landau theory. Therefore, the results 
Eq. (\ref{fractional-charge}) and Eq. (\ref{fractional value}) would 
be valid for realistic systems such as superconducting $Sr_{2}RuO_{4}$. 
A detailed discussion is given in Ref. \cite{Goryo}. 
Evidence that the vortex we consider is a candidate of anyon 
can be seen in the fact that vortex has fractional 
angular momentum.\cite{Wilczek} 
It is expected that transmutation of the statistics occurs.  
The further discussion is needed 
to see the fractional statistics of these vortices experimentally. 
By using a symple Ansatz,  
we found a spatial dependence of a radial electric field numerically. 
The electric field exists around the core 
in the region $r < \lambda$ ($\lambda$ London penetration depth) 
and its magnitude is about $1$ Volt/meter.

In summary, we have discussed a vortex in chiral $p$-wave superconductor 
, where $U(1)$, P- and T-symmetry are broken simultaneously.
The realization of such a superconductivity is argued in $Sr_{2}RuO_{4}$.   
We have investigated the vortex based on the Ginzburg-Landau Lagrangian.  
The Ginzburg-Landau Lagrangian of the system contains 
the Chern-Simons term.   
We have found that the vortex has fractional charge $-\frac{n e}{4}$ and 
fractional angular momentum $-\frac{n^{2}}{16}$ 
which come from the existence of the Chern-Simons term, 
{\it i.e.} P- and T-violation of the ground state. These values are 
topologically stable and do not depend on the form of Ansatz and 
the phenomenological parameters in the Ginzburg-Landau Lagrangian. 
Following the discussion in Ref.\cite{Wilczek}, the vortex 
could obey the fractional statistics. 
We have also investigated the electromagnetic property of the 
vortex. We have found that the electric field is induced by 
the Chern-Simons term near the vortex core. 
It is interesting if these exotic feature is 
observed experimentally. 

Our discussion is valid for other chiral superconductors, 
such as ``anisotropic chiral $p$-wave'' superconductor in which 
the symmetry of the wave function of Cooper pair is 
$(\sin p_{x} + i \sin p_{y})$-wave \cite{Miyake-Narikiyo} 
and $d_{x^{2}-y^{2}}+id_{xy}$-wave superconductor, 
since it has been shown that the Chern-Simons term 
is also induced in the Ginzburg-Landau 
Lagrangian of these superconductors.\cite{Goryo-Ishikawa} 

The author is grateful to K. Ishikawa, N. Maeda,  
Y. Maeno, M. Matsumoto, Y. Okuno, M. Sigrist, and G.E. Volovik  
for useful discussions.

\eject

\end{document}